\documentclass[preprint,twocolumn]{revtex4}

\usepackage{graphicx}
\usepackage{dcolumn}
\usepackage{bm}

\begin{document}

\title{Lifetime of the superconductive state in short and long Josephson
junctions}
\author{\large Giuseppe Augello\footnote{e-mail: augello@gip.dft.unipa.it}, Davide Valenti\footnote{e-mail: valentid@gip.dft.unipa.it},
Andrey L. Pankratov\footnote{e-mail: alp@ipm.sci-nnov.ru}, Bernardo
Spagnolo\footnote{e-mail: spagnolo@unipa.it}}
\affiliation{Dipartimento di Fisica e Tecnologie Relative, Group of
Interdisciplinary Physics\footnote{Electronic address:
http://gip.dft.unipa.it}, Universit\`a di Palermo and CNISM-INFM\\
Viale delle Scienze, I-90128 Palermo, Italy\\
Institute for Physics of Microstructures of RAS, GSP-105, 603950
Nizhny Novgorod, Russia}
\begin{abstract}We study the transient statistical properties of
short and long Josephson junctions under the influence of thermal
and correlated fluctuations. In particular, we investigate the
lifetime of the superconductive metastable state finding the
presence of noise induced phenomena. For short Josephson junctions
we investigate the lifetime as a function both of the frequency of
the current driving signal and the noise intensity and we find how
these noise-induced effects are modified by the presence of a
correlated noise source. For long Josephson junctions we integrate
numerically the sine-Gordon equation calculating the lifetime as a
function of the length of the junction both for inhomogeneous and
homogeneous bias current distributions. We obtain a nonmonotonic
behavior of the lifetime as a function of the frequency of the
current driving signal and the correlation time of the noise.
Moreover we find two maxima in the nonmonotonic behaviour of the
mean escape time as a function of the correlated noise intensity.
\end{abstract}

\maketitle

\section{Introduction}
\label{intro} In recent years a great attention was paid to the
study of Josephson junctions (JJs) because of their use both as
superconducting quantum bits~\cite{Wendin07,Kim06,Zorin06,Berkley03}
and nanoscale superconducting quantum interference devices for
detecting weak magnetic flux change~\cite{Wu08}. This widely used
device is very sensitive to magnetic flux change and it is composed
by two coupling high-T$_c$ JJs in a superconducting ring. JJs are
good candidates to realize superconducting quantum bits (qubits) for
quantum information processing. In particular, they were studied at
very low temperature in devices making use of
charge~\cite{Nakamura99}, flux~\cite{Friedman00} and phase
qubits~\cite{Martinis02}. In JJs, working both at high and low
temperatures, the environment affects strongly the behavior of the
system. In high temperature superconductors (HTSs) the presence of
low-frequency noise, whose intensity is related to the fluctuations
in the bias current, temperature and magnetic field, was
experimentally found~\cite{Hao97}. Also in the low temperature
superconductive devices it is very difficult to avoid the influence
of environment that constitutes mainly a decoherence source for the
system. In particular, the effects on the coherence time of weak
current noise in Josephson vortex qubits (JVQs), composed by an
uniform long JJ, were investigated~\cite{Kim06,Xu05}.

In this framework the study of transient dynamics of JJs in the
presence of noise sources is very interesting for the understanding
of the interaction between these systems and environment. In
particular, the effects of noise strongly influence the
current-voltage characteristic of JJs~\cite{Marx95,Falco74,Pelt06}.
The dynamics of a JJ is studied considering a fictitious Brownian
particle moving in a washboard potential~\cite{Yu02} and the
behavior of the current-voltage characteristic of a JJ is strictly
related to the lifetime of the superconductive metastable state of
the particle. The decay of the particle from the metastable state,
in fact, depends on the fluctuations of the voltage across the
junction. Recently noise induced effects were experimentally
observed in underdamped Josephson junctions~\cite{Yu03,Sun07}, and
the switching to resistive state of an annular Josephson junction
due to thermal activation was analyzed~\cite{Guulevich06}.

 In the present work, we study the transient dynamics of short
overdamped and long JJs under the influence of fluctuating bias
current and oscillating potential. We numerically calculate the
lifetime of the superconductive metastable state also called the
mean switching time (MST) to the resistive state for short JJ (SJJ)
and long JJ (LJJ). We demonstrate the presence of noise induced
effects such as resonant activation
(RA)~\cite{Doe92,Dubkov04,Pank04} and noise enhanced stability
(NES)~\cite{Pank04,Man96,Spa04}. We analyze both the effects of
thermal and correlated noise sources. In SJJ and LJJ we consider
white noise, accounting for the thermal fluctuations, and correlated
(colored) noise separately. Moreover, in LJJ we present an analysis
considering together the effects of white and colored noise. For
given values of frequency of the driving signal and suitable noise
intensity, we find maxima of the lifetime of the superconductive
state. This is an interesting feature for the study of the coherence
time of these devices. Our results hold for low temperature
superconducting devices when we consider only the effects of the
colored noise, and they can be extended to high temperatures, when
both colored and white noise come into play.

\section{Short Overdamped junctions}

\subsection{Model}
\label{sec:1}

The study of SJJs is performed in the framework of the resistively
shunted junction (RSJ) model formalism~\cite{Barone82}. To take into
account the transient dynamics of the system, a fluctuating current
term in the RSJ model equation is considered. We obtain the
following Langevin equation~\cite{Gordeeva08}
\begin{equation}
\frac{d\phi}{dt}=-\omega_c\frac{dU(\phi)}{d\phi}-\omega_c\zeta(t),\label{model}
\end{equation}
where $\phi$ is the order parameter of the system, that is the phase
difference of the wave functions in the ground state between left
and right superconductive sides of the junction. The characteristic
frequency of the Josephson junction is
\emph{$\omega_c$=2e$R_N$$I_c$/$\hbar$}, where \emph{e} is the
electron charge, \emph{$R_N^{-1}$} is the normal conductivity,
\emph{$I_c$} is the critical current and \emph{$\hbar$=h/2$\pi$}
with \emph{h} the Plank constant. In Eq.~(\ref{model}) the time is
normalized to the inverse of the characteristic frequency of the
junction $\omega_c$. In our analysis \emph{$\zeta$(t)} is a colored
noise source, generated using an Ornstein-Uhlenbeck (OU)
process~\cite{Gar04}, characterized by a correlation time $\tau_c$.
The potential profile \emph{U($\phi$)} of Eq.~(\ref{model}) is given
by
\begin{equation}
U(\phi)=1-cos\phi-i(t)\phi,
\end{equation}
where \emph{$i(t)=i_0+f(t)$}, \emph{$i_0=i_b/I_c$} is the constant
dimensionless bias current and \emph{f(t)=Asin$\omega$t} is the
driving current with dimensionless amplitude \emph{$A=i_s/I_c$} and
frequency $\omega$ (\emph{$i_b$} and \emph{$i_s$} represent the bias
current and the driving current amplitude respectively).

The dynamics of a JJ described by Eq.~(\ref{model}) is equivalent to
the dynamics of a particle of coordinate \emph{$\phi$} moving in the
washboard potential \emph{U($\phi$)}. The motion is that of a
Brownian particle because of the presence of the noise
term~\cite{Barone82}. The OU process of Eq.~(\ref{model}) is
represented by the stochastic differential equation~\cite{Gar04}
\begin{equation}
d\zeta(t)=-\frac{1}{\tau_c}\zeta(t)dt+\frac{\sqrt{\gamma}}{\tau_c}dW(t)
\end{equation}
where \emph{$\gamma$} is the noise intensity and \emph{W(t)} is the
Wiener process with the usual statistical properties:
$\left<dW(t)\right> = 0$, and $\left<dW(t)dW(t')\right> =
\delta(t-t')dt$. The correlation function of the OU process is
\begin{equation}
\langle\zeta(t)\zeta(t')\rangle=\frac{\gamma}{2\tau_c}e^{-\frac{{|t-t'|}}{\tau_c}}.
\label{SJJcorrfunct}
\end{equation}
We investigate the dynamics of the Brownian particle in the presence
of a time-dependent nonlinear periodic potential, by solving
numerically Eq.~(\ref{model}), for $\omega_c$=1.

To study the lifetime of the superconductive state we take, as
initial condition for the particle, the minimum of the potential
profile, corresponding to the condition
\emph{$\phi_0$=arcsin($i_0$)} and we calculate the time spent by the
particle to reach the next maximum. We perform a number of
simulations ranging from $N = 5000$ to $N = 10000$ to estimate MST
with a good approximation. The procedure is repeated for different
values of the system parameters such as the amplitude \emph{A} and
frequency \emph{$\omega$} of the driving current signal, the
correlation time \emph{$\tau_c$} and the intensity \emph{$\gamma$}
of the colored noise.

\subsection{Lifetime of metastable state}
\label{sec:2}

Our analysis of the lifetime of the superconducting state concerns
the behavior of the curves representing MST and the corresponding
standard deviation (SD) vs $\omega$, and MST vs $\gamma$, for
different values of $\tau_c$. In the following figures we present
the curves of MST calculated in the presence of colored noise,
including also the curves obtained in the presence of white noise
(see also Ref.~\cite{Gordeeva08}).

In Fig. 1 we report the behavior of MST vs $\omega$, for two values
of the noise intensity, namely $\gamma=0.02$ and $\gamma=0.5$, and
different values of $\tau_c$. The non-monotonic behavior of the
curves shows that the RA phenomenon, already found in the presence
of white noise~\cite{Pank04,Gordeeva08}, appears also with colored
noise. The values of MST around the minimum are influenced by the
variation of $\tau_c$, more strongly for higher values of the noise
intensity.
\begin{figure}[h]
\resizebox{0.95\columnwidth}{!}{
\includegraphics{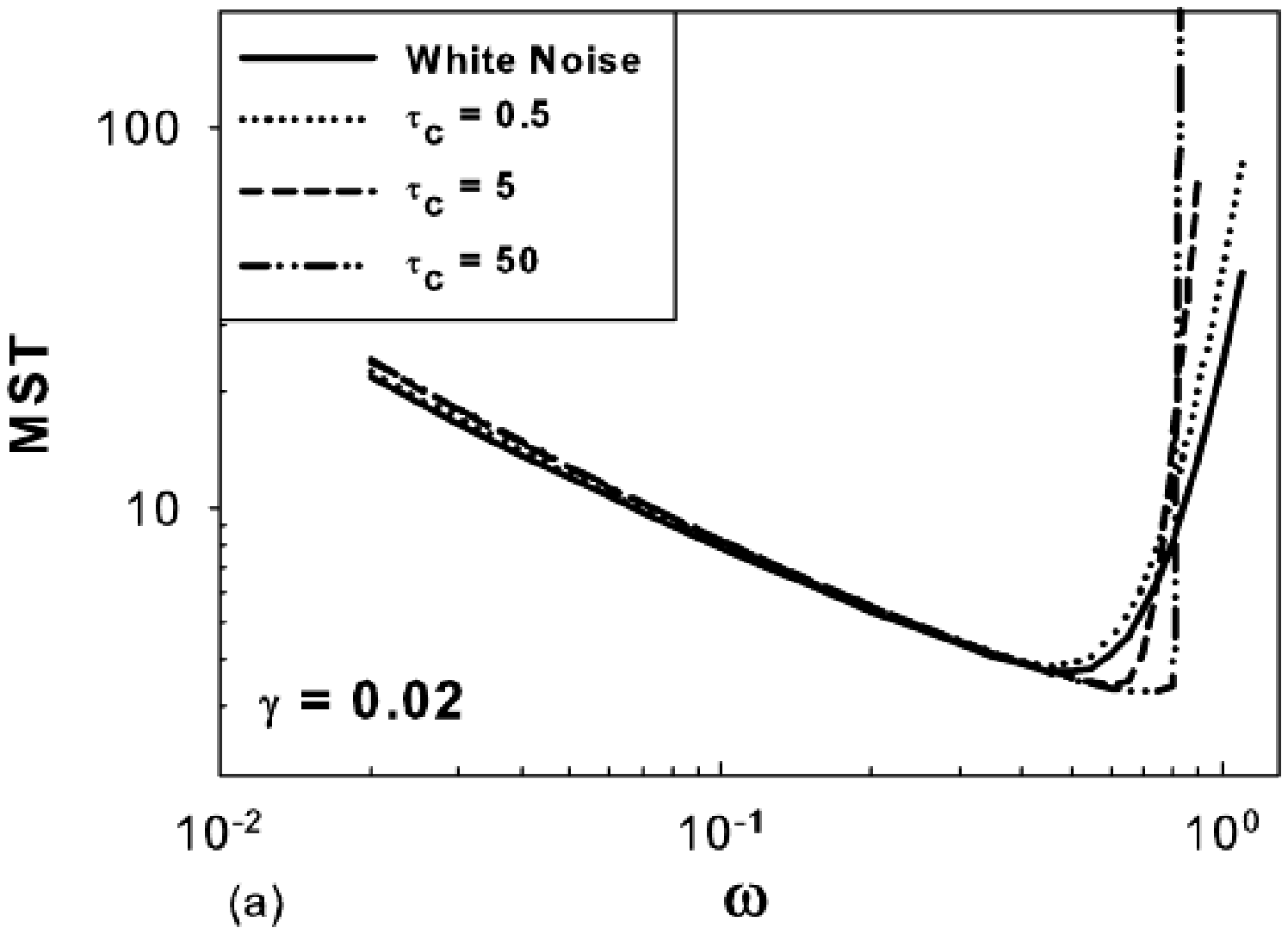}
} \resizebox{0.95\columnwidth}{!}{
\includegraphics{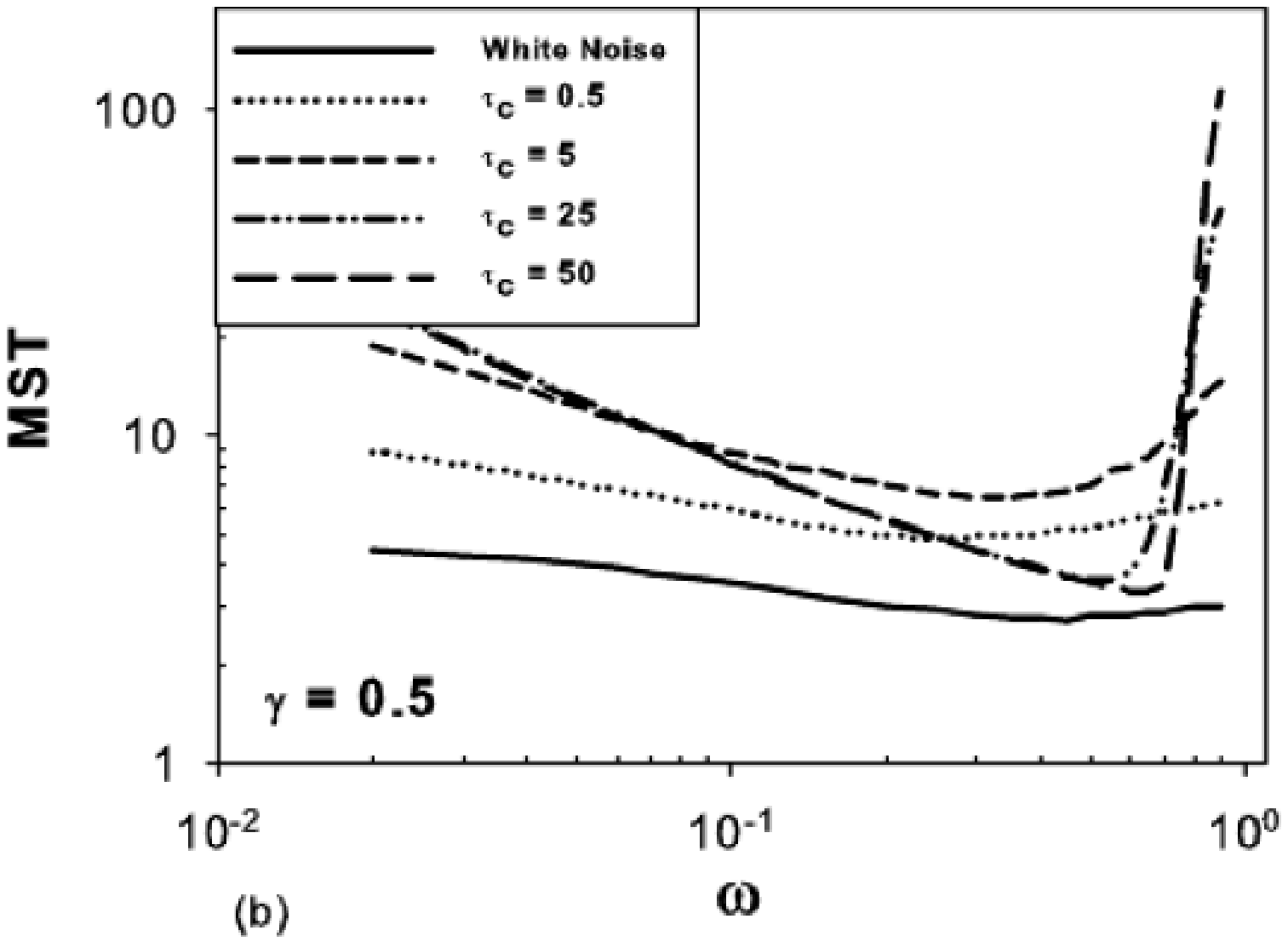}
} \vspace*{8pt} \caption{(a): MST vs {$\omega$} for white noise and
different {$\tau_c$}, {$\gamma$}=0.02. (b):
 MST vs {$\omega$} for white noise and different
{$\tau_c$}, {$\gamma$}=0.5. In both panels $i_0$=0.8 and
A=0.7.}
\label{fig:1}
\end{figure}
Moreover, for higher intensity values (see Fig.~\ref{fig:1}), we
find, in a wide range of frequency (0.3 $<$ $\omega$ $<$ 0.8), a
non-monotonic behavior of MST as a function of the correlation time.
This is  shown more clearly in Fig.~\ref{fig:2}, where we report MST
vs $\tau_c$ for $\omega=0.7$.
\begin{figure}[h]
\resizebox{0.95\columnwidth}{!}{
\includegraphics{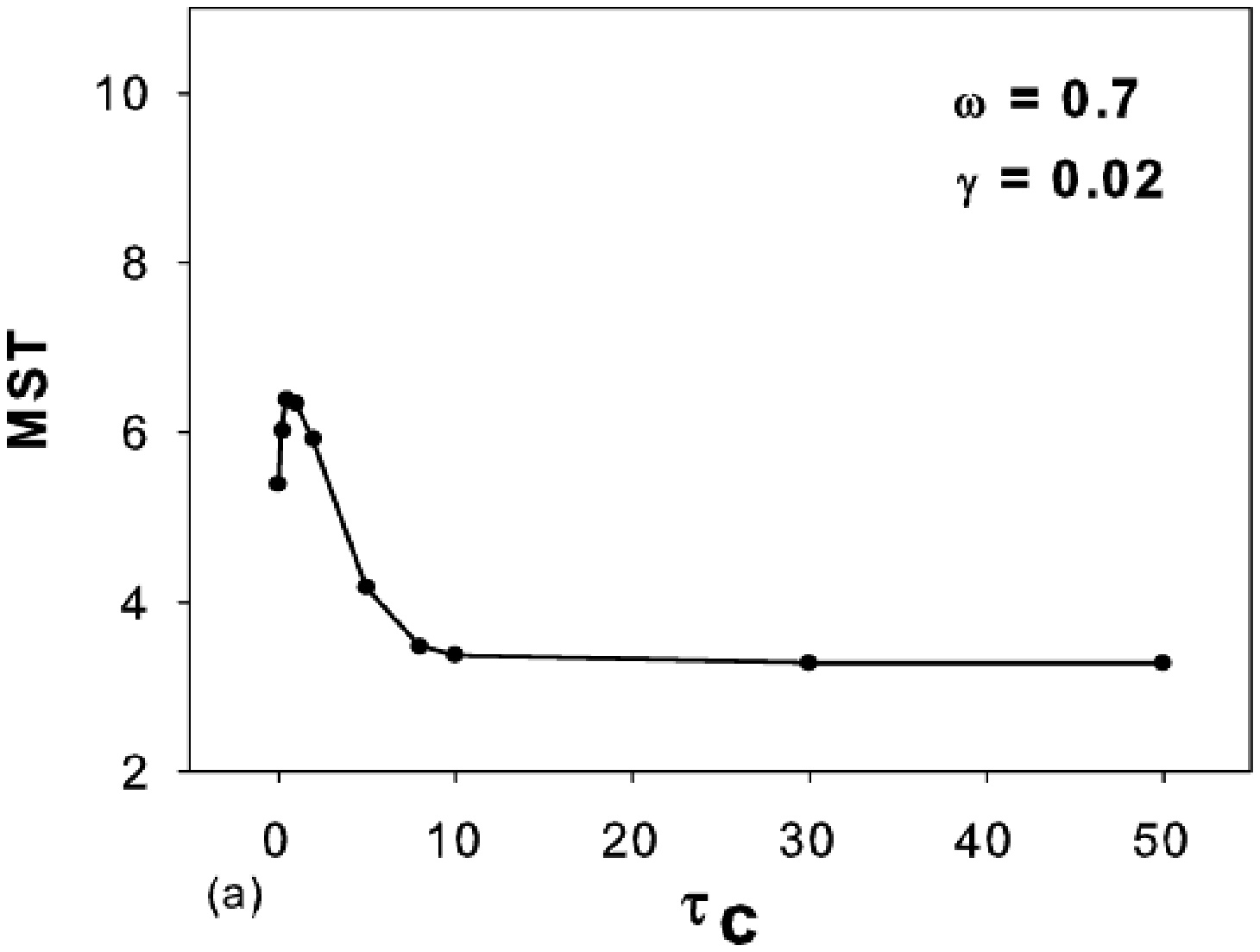}
} \resizebox{0.95\columnwidth}{!}{
\includegraphics{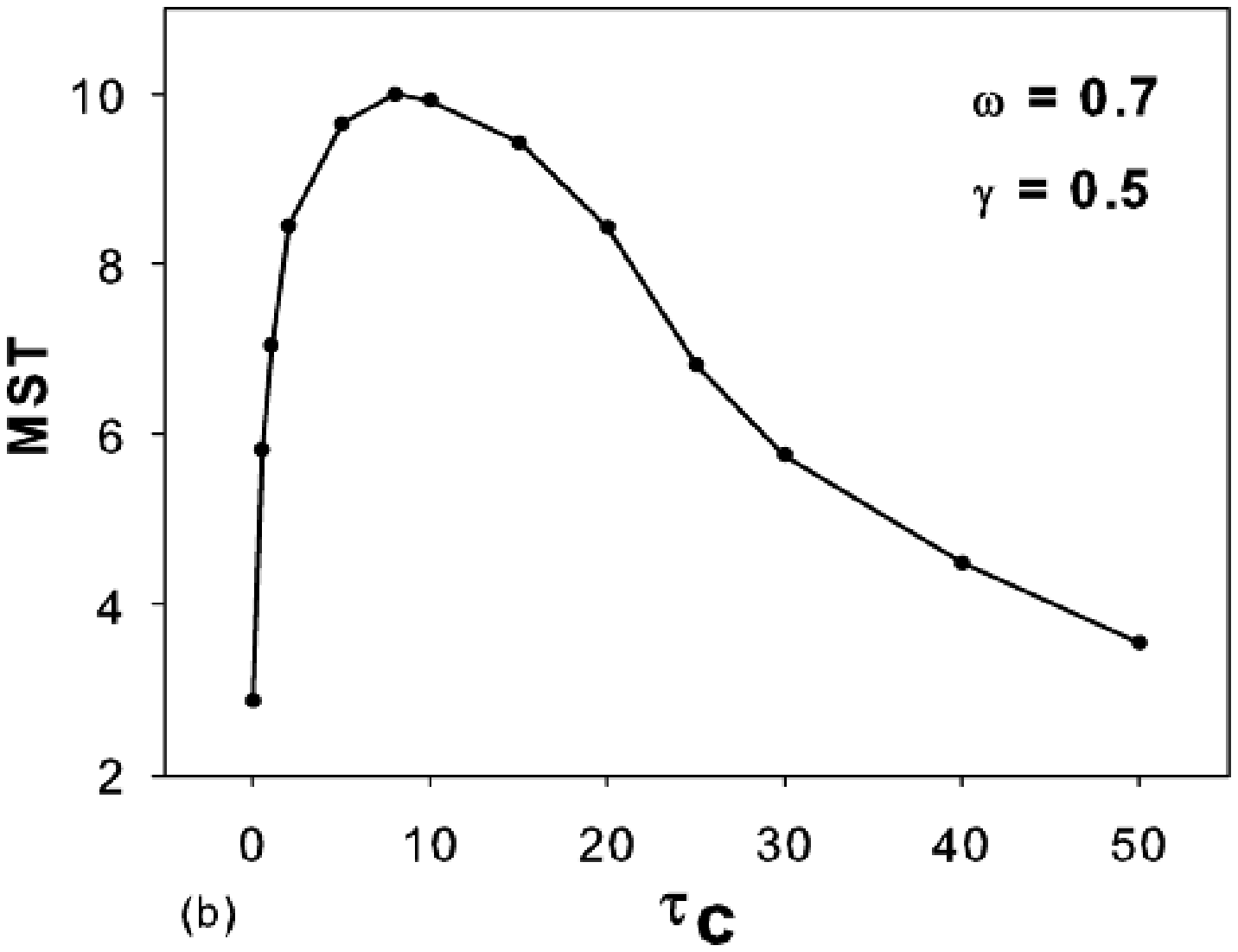}
} \vspace*{8pt} \caption{(a): MST vs {$\tau_c$} for {$\gamma$}=0.02.
(b): MST vs {$\tau_c$} for {$\gamma$}=0.5. In both panels
{$\omega$=0.7}, $i_0$=0.8 and A=0.7.}
\label{fig:2}
\end{figure}
In Fig.~\ref{fig:3} we report the curves of MST and SD vs $\omega$.
We note a range of frequency (0.2 $<$ $\omega$ $<$ 0.8) in which MST
and SD present a minimum, which indicates the suppression of timing
error effects, already found in the presence of white
noise~\cite{Pank04}.
\begin{figure}[h]
\resizebox{0.95\columnwidth}{!}{
\includegraphics{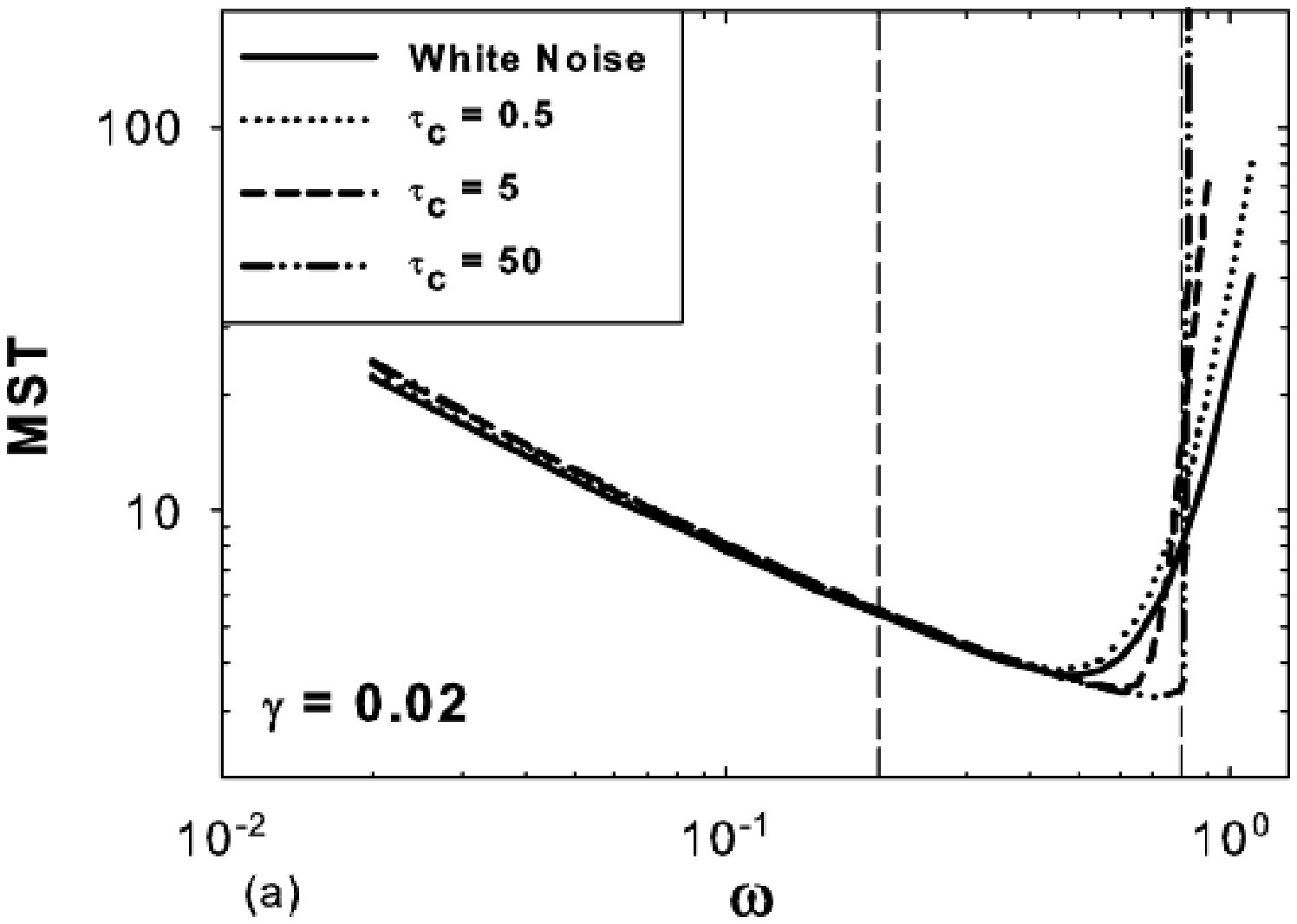}
} \resizebox{0.95\columnwidth}{!}{
\includegraphics{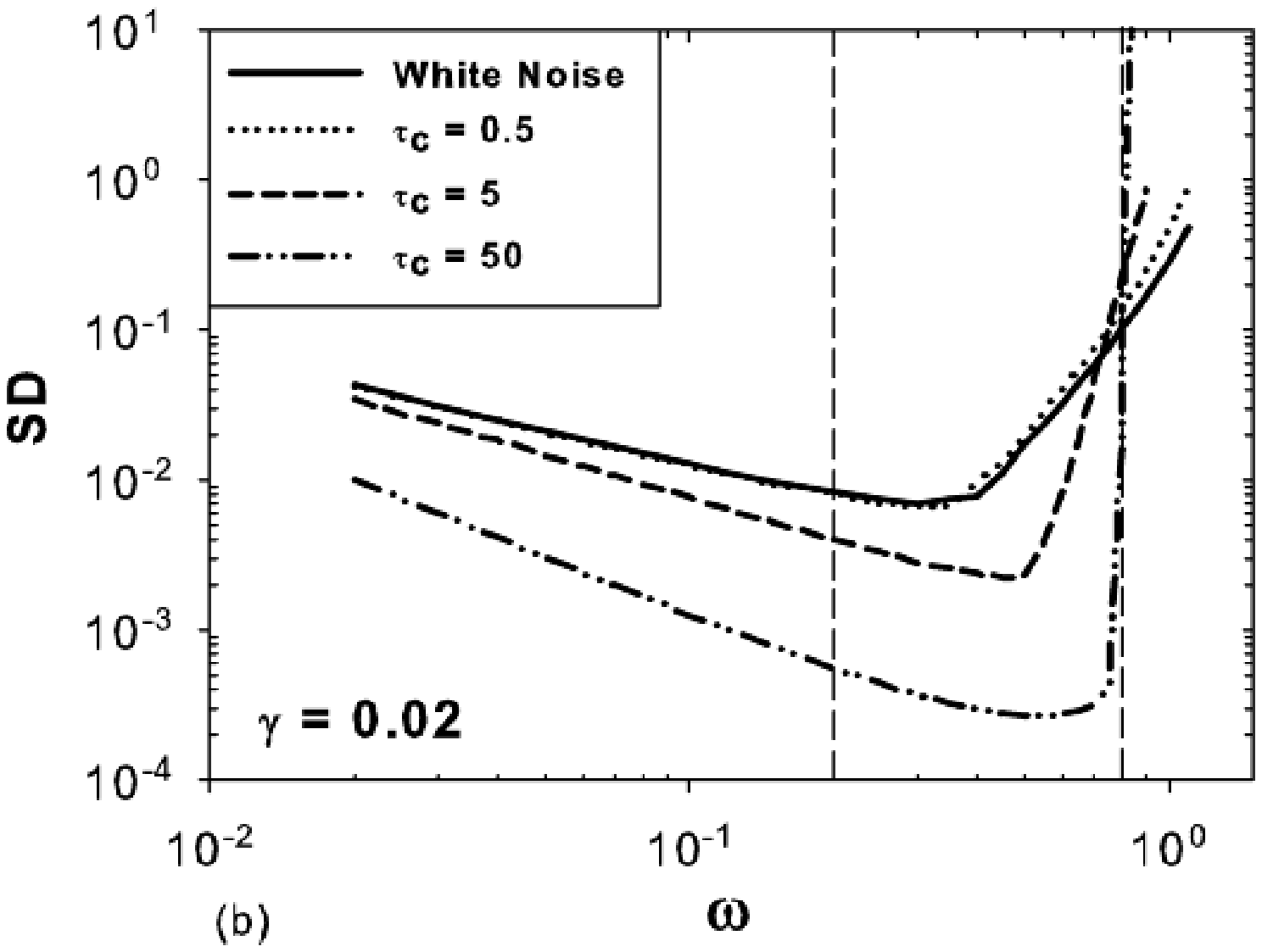}
} \vspace*{8pt} \caption{(a): MST vs {$\omega$} for different
{$\tau_c$}, {$\gamma$}=0.02, $i_0$=0.8 and A=0.7. (b): Standard
deviation (SD) vs {$\omega$}.} \label{fig:3}
\end{figure}
Moreover when we consider the RA phenomenon in the presence of
colored noise, we find a scaling effect depending on the values of
the correlation time. In Eq.~(\ref{SJJcorrfunct}) if the correlation
time
 is greater than the characteristic time scale of the system,
$\tau_c$$\gg$$|$t-t$'$$|$, we obtain
$\langle$$\zeta$(t)$\zeta$(t$'$)$\rangle$
$\approx${$\gamma_{colored}$}/{2$\tau_c$}. By comparison with the
correlation function of the white noise,
$\langle$$\xi$(t)$\xi$(t$'$)$\rangle$={$\gamma_{white}$}$\delta(t-t')$,
we can define the effective intensity of the colored noise scaled by
a factor 1/2$\tau_c$ and equivalent, in this approximation, to the
intensity of the white noise
\begin{equation}
\gamma_{white}=\frac{\gamma_{colored}}{2\tau_c}.
\end{equation}

\begin{figure}[h]
\resizebox{0.95\columnwidth}{!}{
\includegraphics{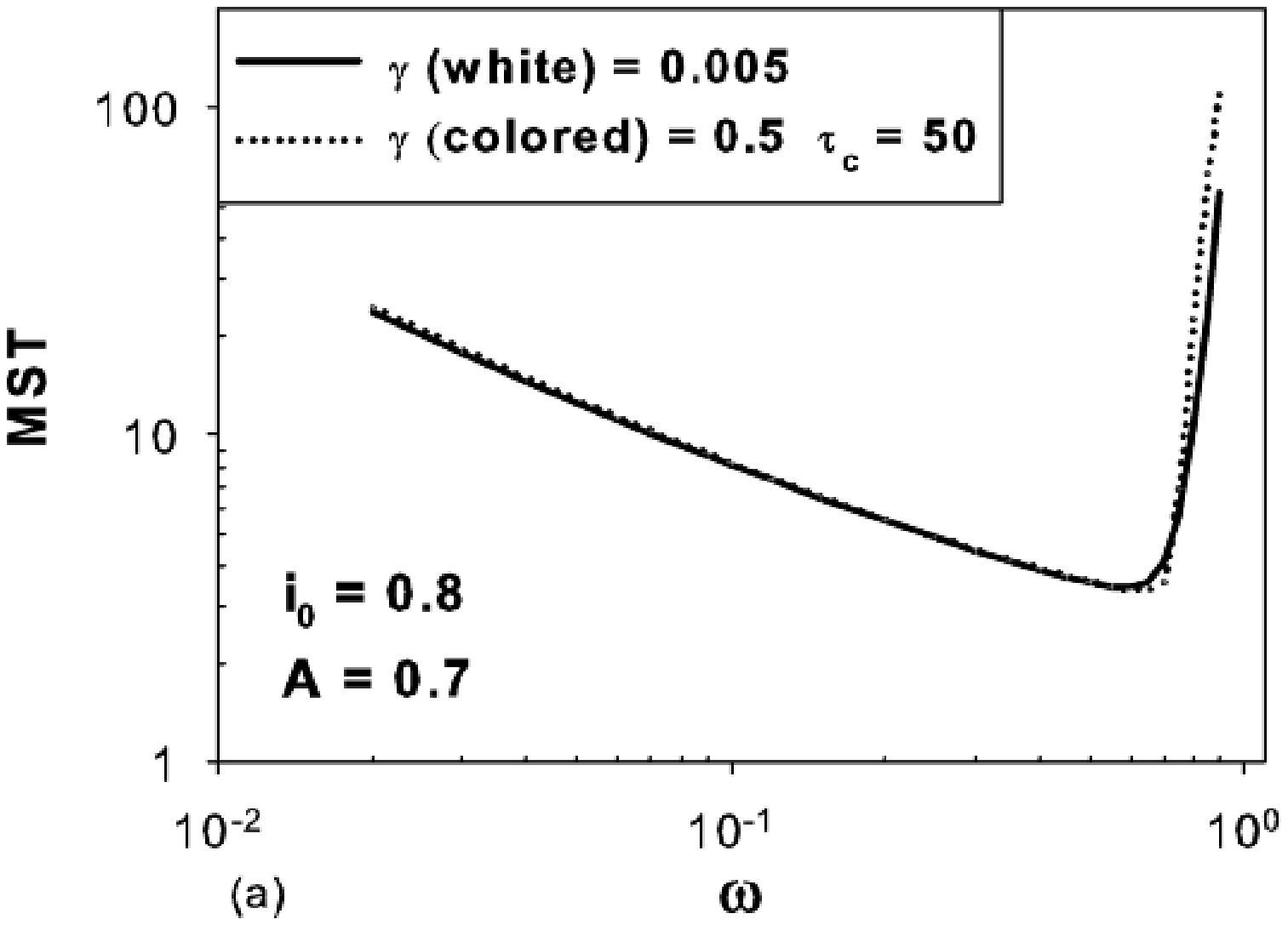}
} \resizebox{0.95\columnwidth}{!}{
\includegraphics{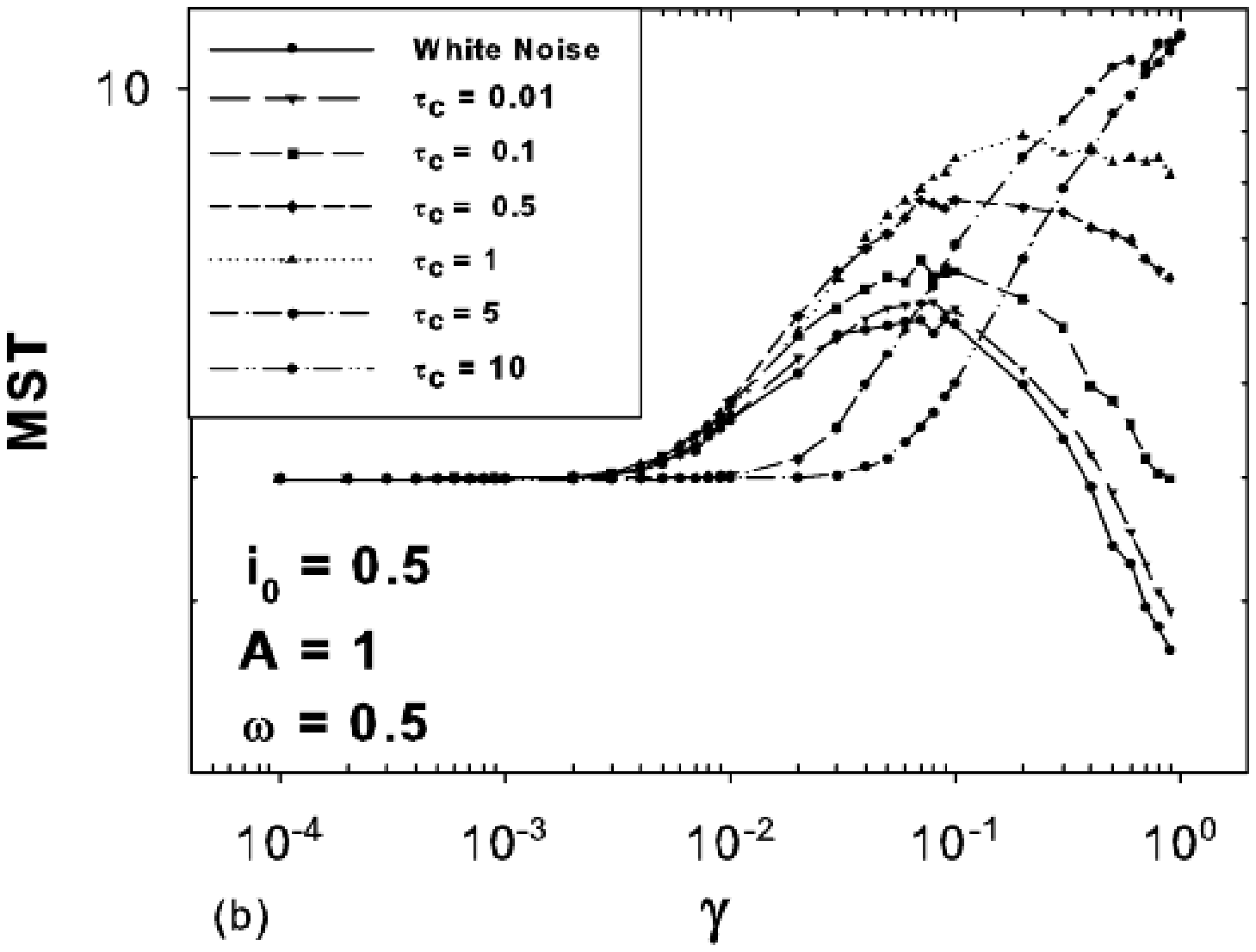}
} \vspace*{8pt} \caption{(a): MST vs $\omega$ with $i_0$=0.8, A=0.7:
white noise (line) and colored noise with $\tau_c$=50 (dots). (b):
MST vs {$\gamma$} for white noise and colored noise with different
values of $\tau_c$, $\omega$=0.5, A=1, $i_0$=0.5.} \label{fig:4}
\end{figure}

We check this equivalence by considering the effect on the system of
a white noise with intensity $\gamma_{white}$=0.005 and a colored
noise with intensity $\gamma_{colored}$=0.5, setting $\tau_c$=50. In
Fig. \ref{fig:4}a, we report the curves of MST vs $\omega$
calculated for these two values of noise intensity. We see that for
MST less than 50 there is a good agreement between the curves. In
these conditions the evolution occurs with time scale smaller than
the correlation time and the system is not affected by the noise
memory.
In  Fig.~\ref{fig:4}b, we show the behavior of MST as a function of
the noise intensity for different values of the correlation time. We
note the appearance of noise enhanced stability (NES), a phenomenon
already found in short JJs in the presence of white
noise~\cite{Pank04,Gordeeva08}. Here we observe a range of values of
correlation time, namely 0.01$<$$\tau_c$$<$1, in which the curves
present a non-monotonic behavior. For $\tau_c$=5,10, the
non-monotonic behavior disappears.

\section{Long junctions}
\subsection{Model}
\label{sec:3}

The investigation of LJJs is developed in the framework of the
sine-Gordon model~\cite{Barone82,Nori08}. The dynamics of a LJJ is
described by the motion of a phase string in a one dimensional
washboard potential. The transient dynamics is represented by a
nonlinear partial differential equation with a stochastic
term~\cite{Fed07,Fed08}
\begin{equation}
\beta\frac{\partial^2\phi}{\partial t^2}+\frac{\partial\phi}{\partial
t}+\frac{\partial^2\phi}{\partial x^2}=i(x)-\sin\phi+i_f(x,t),
\label{SGmodel}
\end{equation}
where $\beta=\omega_c RC$, with $R$ and $C$ the LJJ equivalent
resistance and capacitance, respectively. Here $i(x)$ and $i_f(x,t)$
are the bias current density and the fluctuating current density
normalized to the critical current density $I_c$ of the junction,
respectively. We treat both the homogeneous bias current density
case, namely $i(x)=i_b$, and the inhomogeneous one, namely
$i(x)=(i_b L)/(\pi\sqrt{x(L-x)})$, where $L=l/\lambda_J$ with $l$
the length of the junction and $\lambda_J$ the Josephson penetration
depth~\cite{Fed07,Fed08}. In Eq.~(\ref{SGmodel}) time is normalized
to the inverse of the characteristic frequency, $\omega_c=2eRI_c/
\hbar$, of the junction. Analogously length is normalized to the
Josephson penetration depth. The boundary conditions for
Eq.~(\ref{SGmodel}) are:
\begin{equation}
\frac{\partial\phi(0,t)}{\partial
x}=\frac{\partial\phi(L,t)}{\partial x}=0 .
\label{SGBoundary}
\end{equation}
As initial position of the phase string in the washboard potential,
we consider the minimum, represented by the condition
$\phi(x,0)=arcsin(i_b)$.

First we present our analysis concerning the effects of a
fluctuating current signal described by a correlated noise. Then we
consider the effects of both colored and white noise on the system,
by inserting in Eq.~(\ref{SGmodel}) a fluctuating term given by the
sum of a correlated noisy current signal, $i_{cn}(x,t)$, and a
thermal current signal, $i_{wn}(x,t)$, namely
\begin{equation}
i_f(x,t)=i_{cn}(x,t)+i_{wn}(x,t). \label{total fluc curr}
\end{equation}

The integration of Eq.~(\ref{SGmodel}) is performed using an
implicit finite-difference method~\cite{Lom04,Zhang91}, that allows
to obtain a system of equations, whose solution is calculated by a
tridiagonal algorithm. The integration step both for the time and
the space is $0.05$. The number of simulation ranges from $N = 100$
to $N = 5000$.

\subsection{Effects of correlated noise}
\label{sec:4}

\begin{figure}[h]
\resizebox{0.95\columnwidth}{!}{
\includegraphics{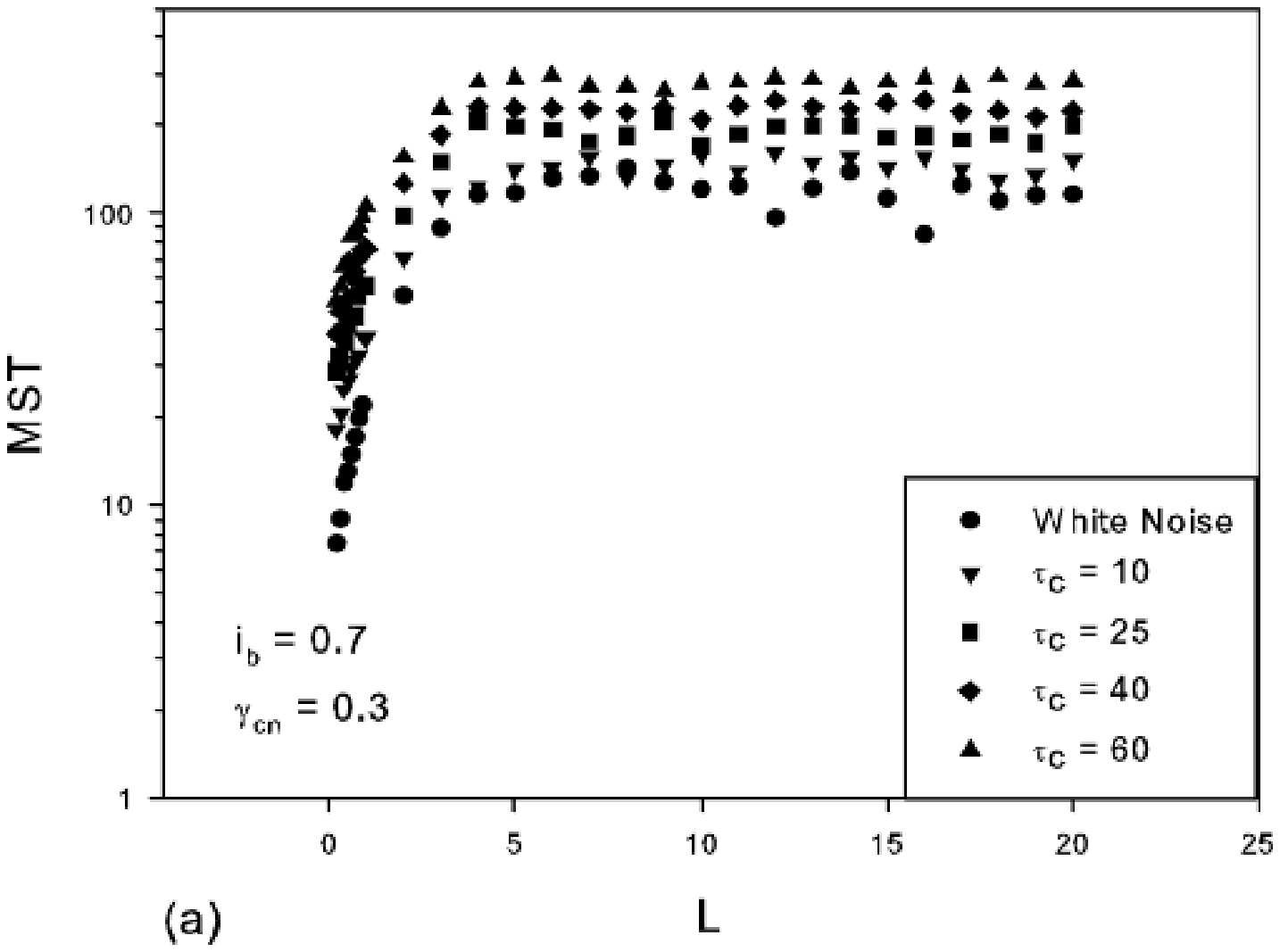}
} \resizebox{0.95\columnwidth}{!}{
\includegraphics{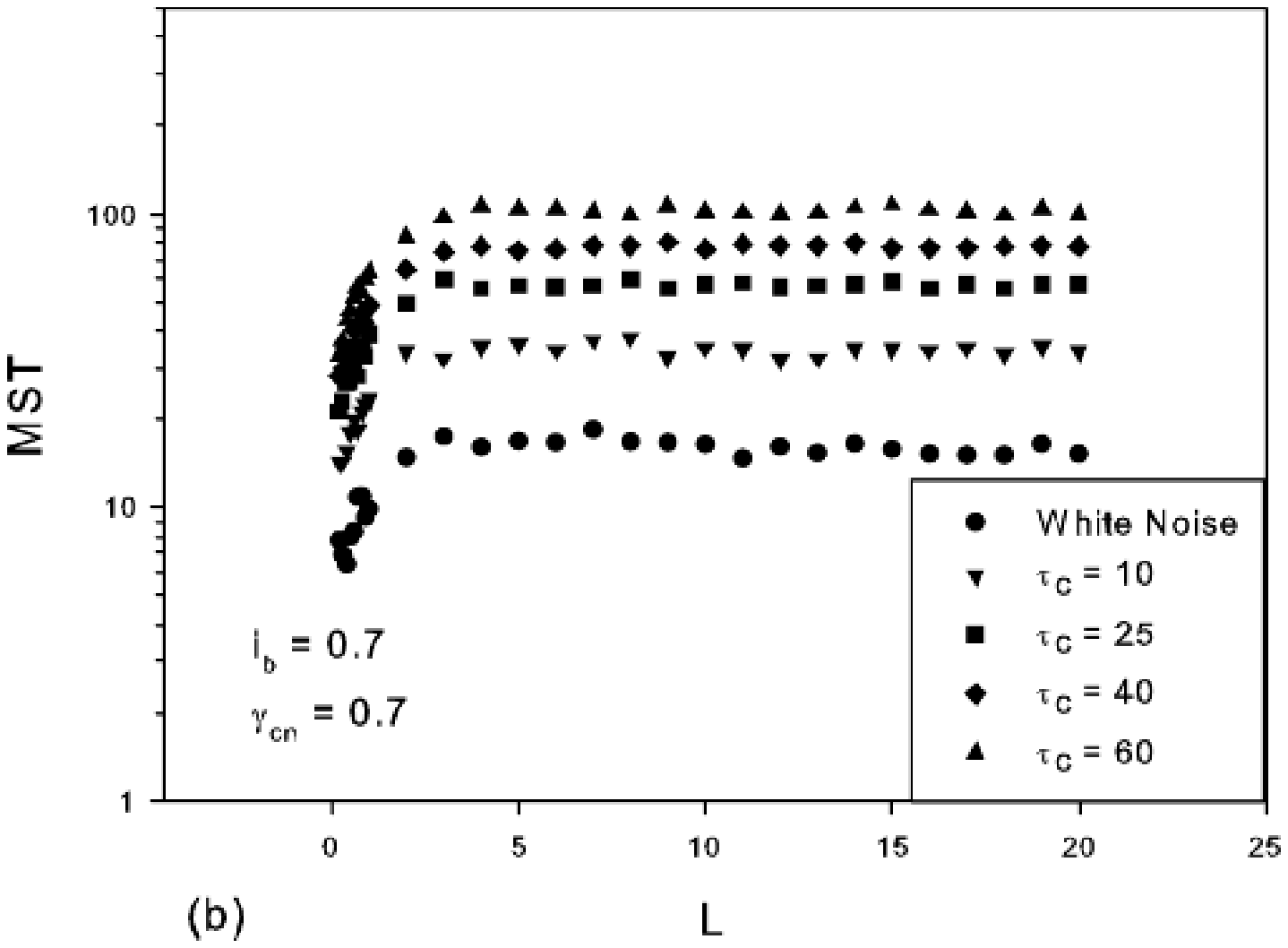}
} \vspace*{8pt} \caption{(a): MST vs L with $i_b$=0.7,
$\gamma_{cn}$=0.3. (b): MST vs L with $i_b$=0.7, $\gamma_{cn}$=0.7.}
\label{fig:5}
\end{figure}

\begin{figure}[h]
\resizebox{0.95\columnwidth}{!}{
\includegraphics{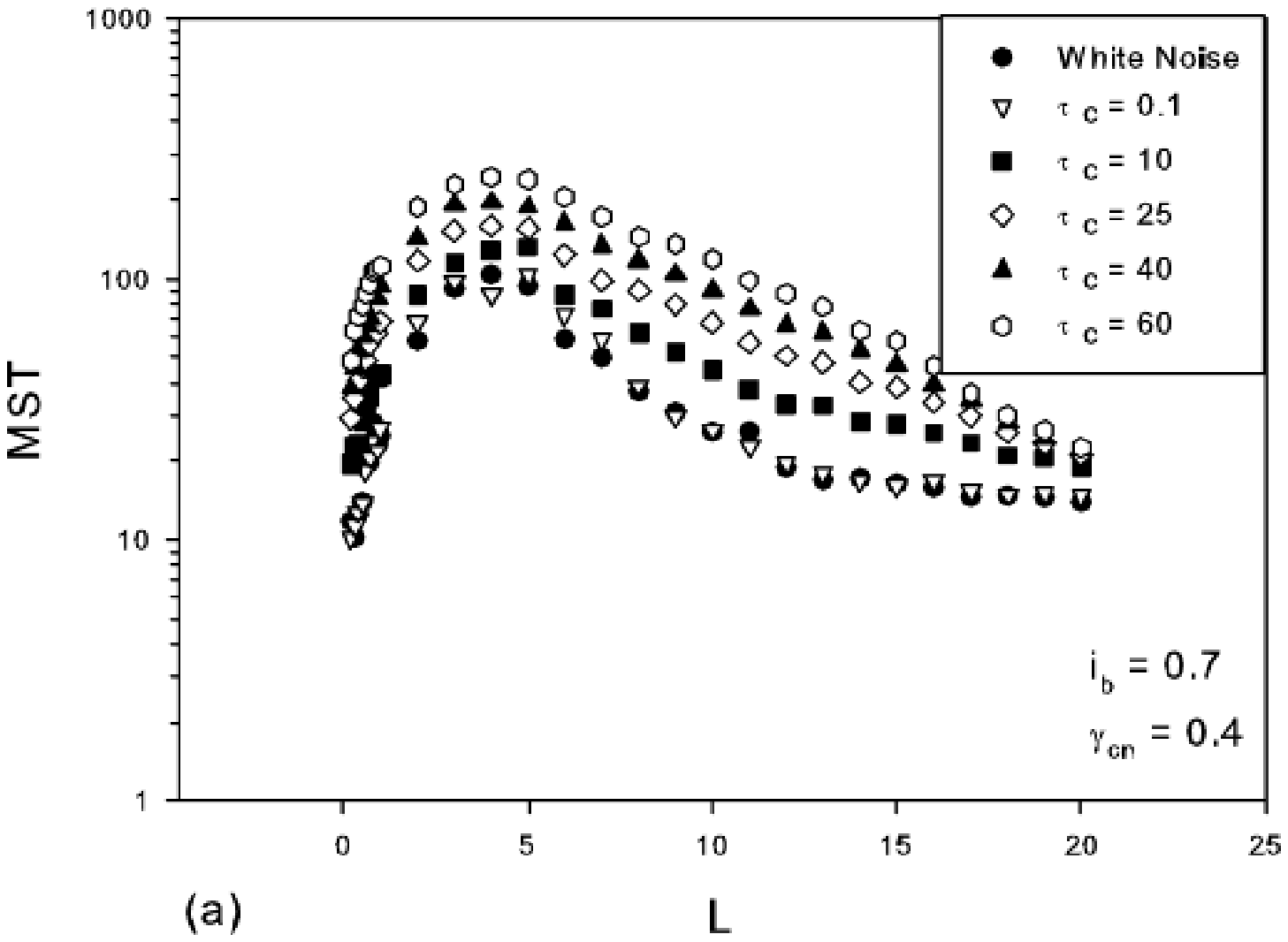}
} \resizebox{0.95\columnwidth}{!}{
\includegraphics{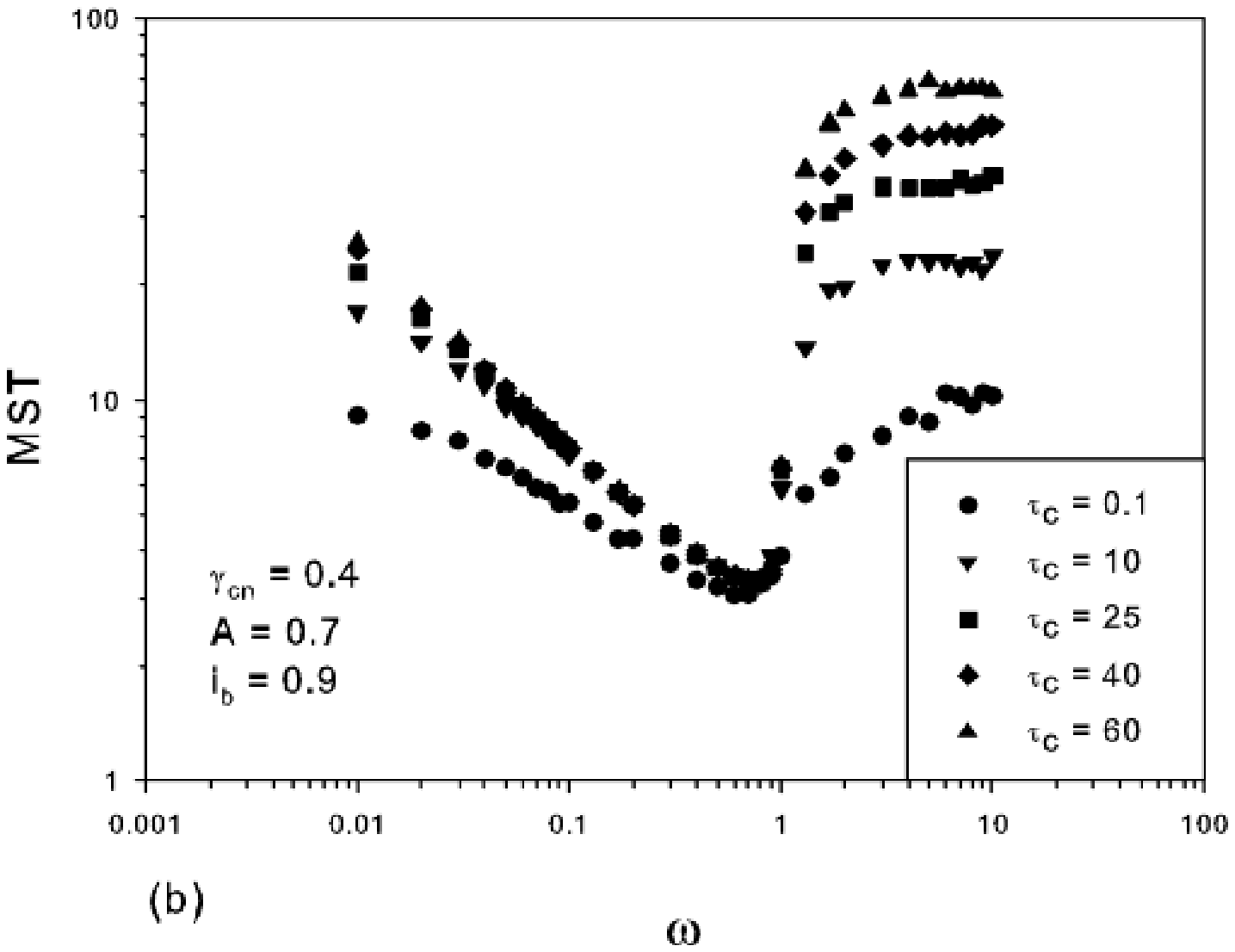}
} \vspace*{8pt} \caption{(a): MST vs L with $i_b$=0.7,
$\gamma_{cn}$=0.4. (b): MST vs $\omega$ with $\gamma_{cn}$=0.4,
A=0.7 and $i_b$=0.9,} \label{fig:6}
\end{figure}

\begin{figure}[h]
\resizebox{0.95\columnwidth}{!}{
\includegraphics{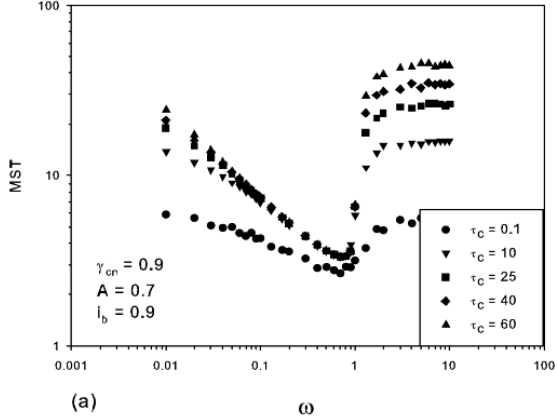}
} \resizebox{0.95\columnwidth}{!}{
\includegraphics{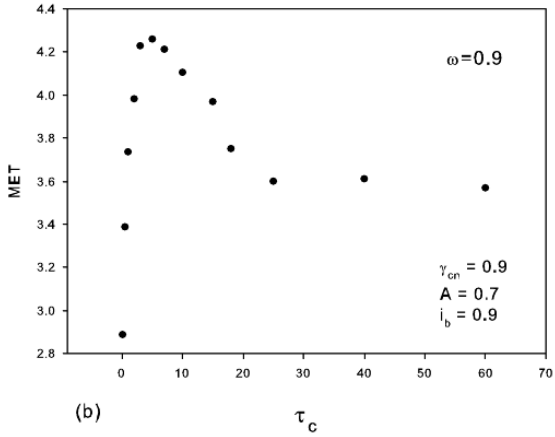}
} \vspace*{8pt} \caption{(a): MST vs $\omega$ with
$\gamma_{cn}$=0.9, A=0.7 and $i_b$=0.9. (b): MST vs $\tau_c$ with
$\gamma_{cn}$=0.9, A=0.7 and $i_b$=0.9.} \label{fig:7}
\end{figure}

\begin{figure}[h]
\resizebox{0.95\columnwidth}{!}{
\includegraphics{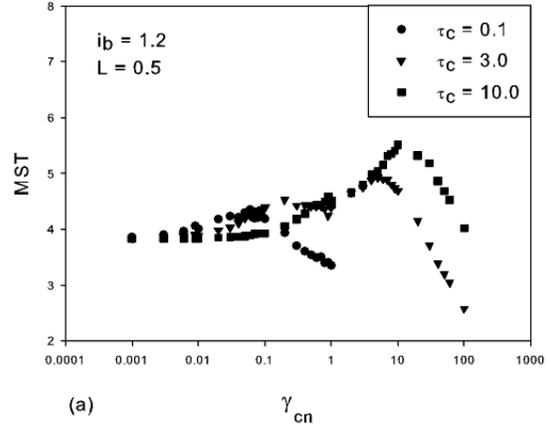}
} \resizebox{0.95\columnwidth}{!}{
\includegraphics{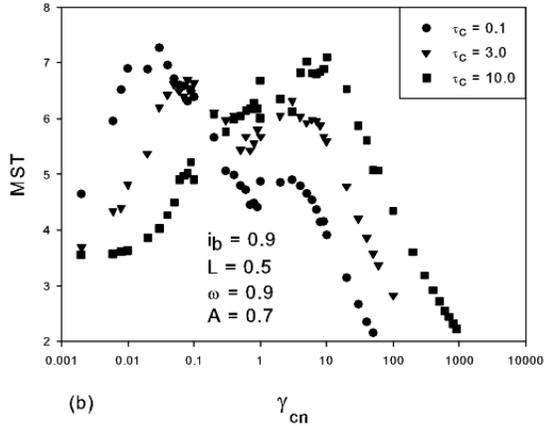}
} \vspace*{8pt} \caption{(a): MST vs $\gamma_{cn}$ in the resistive
state ($i_b=1.2$) with L=0.5 (b): MST vs $\gamma_{cn}$ in the
superconductive state ($i_b=0.9$) with L=0.5, $\omega=0.9$ and
A=0.7.} \label{fig:8}
\end{figure}

\begin{figure}[h]
\resizebox{0.95\columnwidth}{!}{
\includegraphics{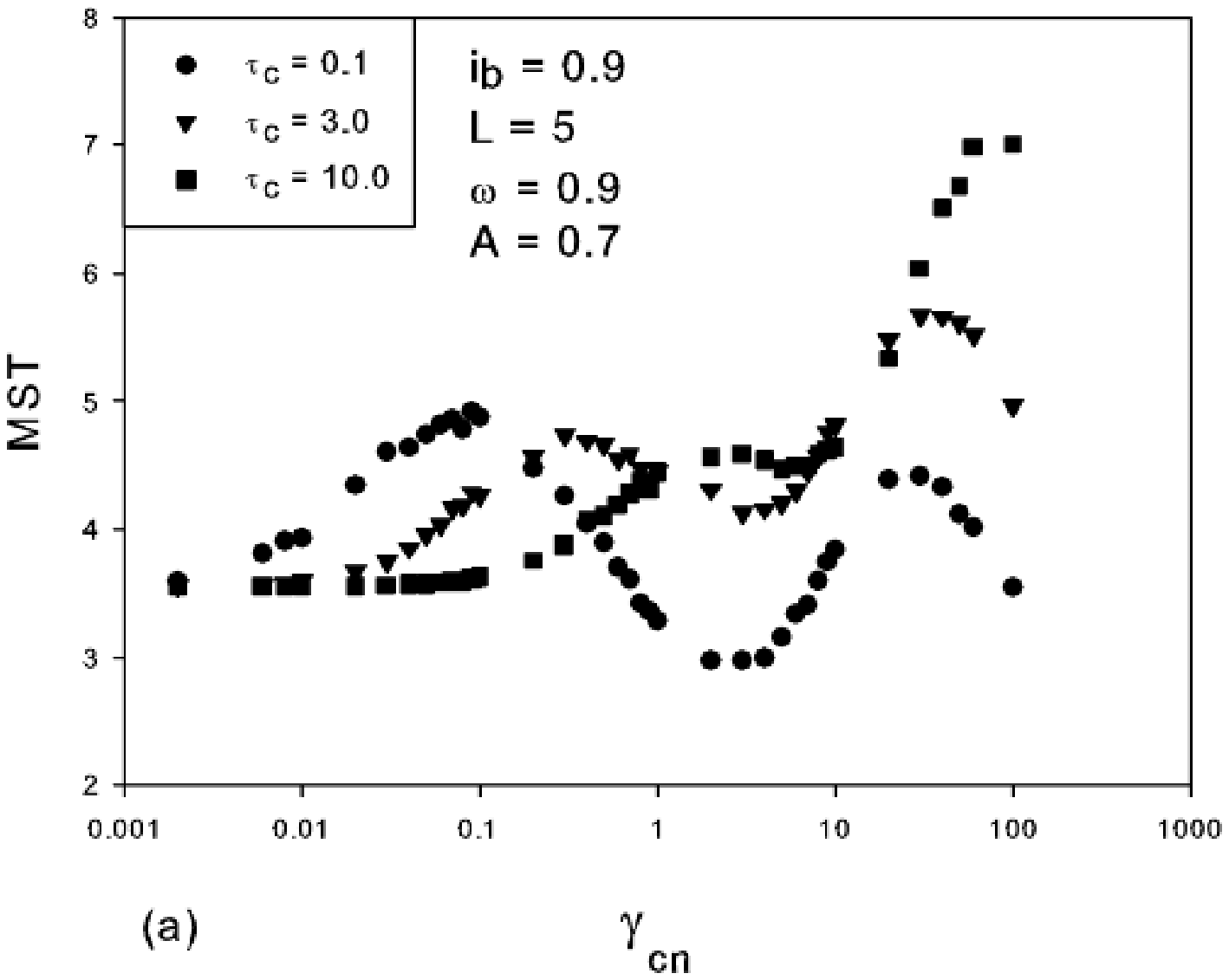}
} \resizebox{0.95\columnwidth}{!}{
\includegraphics{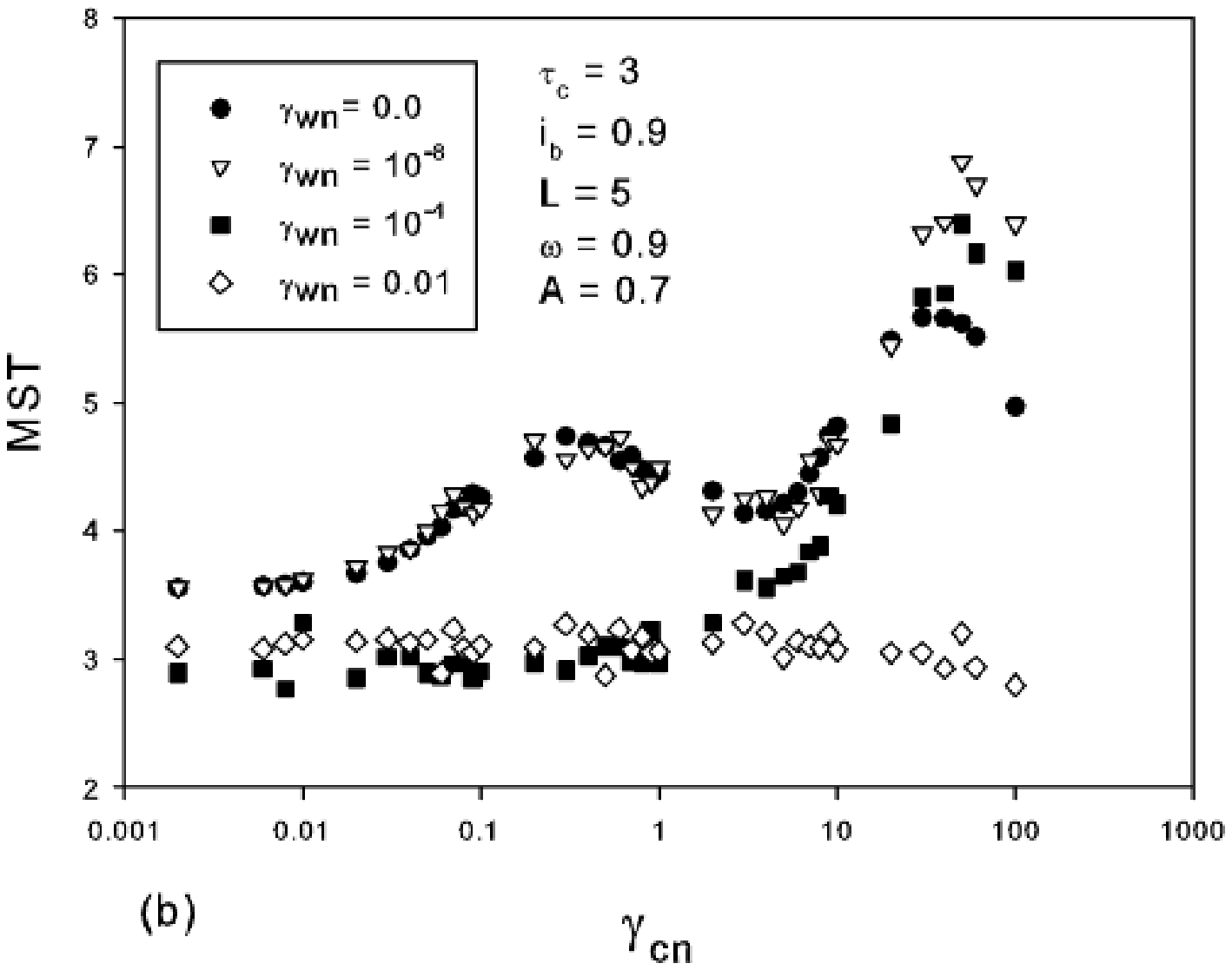}
} \vspace*{8pt} \caption{(a): MST vs $\gamma_{cn}$ in the
superconductive state ($i_b=0.9$) with L=5, for different values of
correlation time. (b): MST vs $\gamma_{cn}$ in the superconductive
state ($i_b=0.9$) with L=5, for different values of white noise
intensity.} \label{fig:9}
\end{figure}

Here we study the effects of a correlated noise on the transient
dynamics of a LJJ, neglecting, in Eq.~(\ref{SGmodel}), the term
$i_{wn}(x,t)$ and considering a fluctuating current signal
$i_{cn}(x,t)$ with correlation function
\begin{equation}
\langle
i_{cn}(x,t)i_{cn}(x',t')\rangle=\frac{(2\gamma_{cn})}{2\tau_c}
\delta(x-x')e^{-\frac{{|t-t'|}}{\tau_c}}, \label{CorrFunctionLJJ}
\end{equation}
where $2\gamma_{cn}$ is the intensity of the correlated noise.
Therefore, the current density $i_{cn}(x,t)$ is subjected to
fluctuations that are correlated in time (colored noise), showing a
$\delta$-correlated behavior in space (white noise). In
Fig.~\ref{fig:5} we report the curves of MST vs the length of the
junction, in the homogeneous bias current case, for different values
of $\tau_c$, including white noise ($\tau_c =
0$)~\cite{Fed07,Fed08}. We find a decrease of MST as the intensity
of the colored noise $\gamma_{cn}$ increases (see Fig.~\ref{fig:5}a
($\gamma_{cn} = 0.3$) and Fig.~\ref{fig:5}b ($\gamma_{cn} = 0.7$)).
Moreover, we observe that, for fixed values of the noise intensity,
MST increases as the correlation time becomes larger.
We also present the numerical results for the case of inhomogeneous
bias current. In Fig.~\ref{fig:6}a we show the behavior of MST as a
function of the dimensionless length \emph{L}. We observe that the
curves present a non-monotonic behavior with a maximum in
correspondence of $L=5$.
In order to investigate the presence of RA phenomenon, in
Eq.~(\ref{SGmodel}) we replace the term $i(x)$ with a homogeneous
oscillating driving current signal given by $i(t)=i_b+Asin(\omega
t)$, with \emph{A} and $\omega$ amplitude and angular frequency,
respectively, of the periodical driving signal. The curves obtained
by numerical simulation are presented in Fig.~\ref{fig:6}b and
Fig.~\ref{fig:7}a. In both figures we note the presence of a minimum
in the curves of MST vs $\omega$. This is the signature of the
resonant activation phenomenon.
Moreover in Fig.~\ref{fig:7}a, in the frequency range
$0.3<\omega<1.0$, we find that curves with different $\tau_c$ are
overlapping. In order to investigate in detail the behavior in this
frequency range we report, for the same parameter values of
Fig.~\ref{fig:7}a, the MST as a function of $\tau_c$ for
$\omega=0.9$, finding a nonmonotonic behavior. We also investigate
the presence of NES for the homogeneous current case in the
resistive state (Fig.~\ref{fig:8}a) without driving signal, and in
the superconductive state in the presence of a driving current
signal (Fig.~\ref{fig:8}b). The curves of Fig.~\ref{fig:8}a present
a maximum for different values of $\tau_c$. In Fig.~\ref{fig:8}b we
find, in all curves, a double peak. The comparison of the two panels
in Fig.~\ref{fig:8} shows that the driving signal causes the
increase of the two maxima (see Fig.~\ref{fig:8}b) already present
in the absence of periodical term (see Fig.~\ref{fig:8}a).
The non-monotonic behavior of MST as a function of the noise
intensity $\gamma_{cn}$ indicates the presence of a NES effect. The
curves of Fig.~\ref{fig:8} have been calculated for $L=0.5$. In
Fig.~\ref{fig:9}a, we report the curves of MST vs $\gamma_{cn}$ for
$L=5$, obtained in the presence of periodical driving signal for the
superconductive state. We note that, using the same noise
intensities, for longer junction the decay time is shorter. This can
be explained recalling that the effect of the spatial correlation
disappears on a longer distance. This corresponds to the presence of
white noise, which is responsible for faster dynamics and, then,
shorter escape time.
\subsection{Effects both of correlated and thermal noise}
\label{sec:5} In this section we investigate the effects both of
thermal and correlated noise, considering in Eq.~(\ref{SGmodel}) the
fluctuating current term $i_f(x,t)$ given by Eq.~(\ref{total fluc
curr}). In Fig.~\ref{fig:9}b, we present the curves of MST vs
$\gamma_{cn}$ for different white noise intensities $\gamma_{wn}$,
with $L=5$.

We note that when $\gamma_{wn}$ is greater than $\gamma_{cn}$ the
effects of colored noise disappear. If we consider $\gamma_{wn}$
suitably lower than $\gamma_{cn}$, the effects of colored noise
become evident. In fact, when $\gamma_{cn}$ is at least five order
of magnitude higher than $\gamma_{wn}$, the curve of MST vs
$\gamma_{cn}$ (empty triangles) matches with good agreement that
(black circles) obtained in the absence of white noise (see Fig.
\ref{fig:9}b for $\gamma_{wn}$=$10^{-8}$).


\section{Conclusions}
\label{sec:6} We studied the effects of white and colored noise on
the transient dynamics of short and long Josephson junctions. We
investigated the lifetime of the metastable state finding noise
induced effects, namely resonant activation and noise enhanced
stability. The results obtained throw light upon the role played by
different noise sources in the dynamics of superconductive devices,
explaining how random fluctuations influence the mean switching time
of short and long Josephson junctions. We found that these times are
strictly affected by the characteristic parameters of the system
such as intensity of noise, frequency of the driving signal and
correlation time of the colored noise.

\end{document}